%% file: main.tex
\documentclass[aps,floatfix,superscriptaddress,twocolumn,tightenlines,amsmathm,textcomp,amssymb,nofootinbib,10pt,nofootinbib]{revtex4-1}
\input{formatting}

\begin{document}
\title{\normalfont\Large{Emulation of satellite up-link quantum communication with entangled photons}}

\author{T. Jaeken$^*$}
\affiliation{Institute of Photonics and Quantum Sciences, School of Engineering and Physical Sciences, Heriot-Watt University, Edinburgh EH14 4AS, UK}

\author{A. Pickston$^*$}
\affiliation{Institute of Photonics and Quantum Sciences, School of Engineering and Physical Sciences, Heriot-Watt University, Edinburgh EH14 4AS, UK}

\author{F. Redza}
\affiliation{Institute of Photonics and Quantum Sciences, School of Engineering and Physical Sciences, Heriot-Watt University, Edinburgh EH14 4AS, UK}

\author{T, Jennewein}
\affiliation{Institute for Quantum Computing and Department of Physics \& Astronomy, University of Waterloo, 200 University Ave W,
Waterloo, N2L 3G1, Ontario, Canada.}
\affiliation{Department of Physics, Simon Fraser University, 8888 University Dr W, Burnaby, V5A 1S6,
British Columbia, Canada.}

\author{A. Fedrizzi}
\affiliation{Institute of Photonics and Quantum Sciences, School of Engineering and Physical Sciences, Heriot-Watt University, Edinburgh EH14 4AS, UK}

\def\thefootnote{*}\footnotetext{These authors contributed equally to this work}\def\thefootnote{\arabic{footnote}}

\begin{abstract}
Quantum communication rates in terrestrial quantum networks are fundamentally limited by fibre loss, even in the presence of quantum repeaters. Satellites offer a solution for long-distance communication, with the most commonly explored scenario involving prepare-and-measure protocols connecting from orbit to a trusted-node ground station via free-space down-links. In contrast, up-link scenarios allow for entanglement to be distributed between a satellite and remote end users in terrestrial networks, eliminating any trust requirement on the ground station.
Here we demonstrate an ultra-bright source of far-non-degenerate entangled photons and perform quantum key distribution in emulated high-loss satellite scenarios. With a loss profile corresponding to that of one of the pioneering Micius up-link experiments, and a terrestrial end user separated by 10~km of telecom fibre we achieve secure key bit accumulation of 5.2~kbit in a single emulated overpass in the asymptotic limit. Our results confirm the viability of upcoming low-Earth orbit receiver satellite missions.
\end{abstract}


\maketitle
One of the main stumbling blocks for globally distributed quantum-enhanced tasks, is the well established bound on direct point-to-point transmission of quantum information~\cite{pirandola_fundamental_2017,takeoka_fundamental_2014}.
Quantum repeaters---untrusted intermediate nodes that perform entanglement swapping and purification---are widely assumed to eventually overcome this problem~\cite{briegelQuantumRepeatersRole1998, azuma_quantum_2023,munroQuantumRepeaters2015,xuSecureQuantumKey2020,muralidharanOptimalArchitecturesLong2016,wehnerQuantumInternetVision2018,loMeasurementDeviceIndependentQuantumKey2012,lucamariniOvercomingRateDistance2018,Pirandola:20}.
However, despite impressive results~\cite{thomas_deterministic_2024,azuma_all-photonic_2015,li_experimental_2019,sangouardQuantumRepeatersBased2011b,tittelPhotonechoQuantumMemory2010,childressFaultTolerantQuantumCommunication2006}, the demanding performance criteria for a fully-fledged repeater node have not yet been met. Even once that is the case, rates in a long-distance repeater network will be low. 

Quantum-enabled satellites have presented themselves as both an alternative and complementary means to exchange quantum resources between distant locations.
By using free space channels between the ground and satellites, the exponential loss scaling in conventional fibre infrastructure is circumvented.
Initial feasibility studies for quantum-enabled satellites~\cite{bonatoFeasibilitySatelliteQuantum2009,hasan_qeyssat_2014,villoresiExperimentalVerificationFeasibility2008,rarityGroundSatelliteSecure2002,wangDirectFullscaleExperimental2013,peng_experimental_2005} were followed up by the Chinese satellite Micius.
Launched in 2016, this pioneering quantum-enabled satellite mission demonstrated satellite-based entanglement distribution over 1200~km~\cite{yinSatellitebasedEntanglementDistribution2017a,lu_micius_2022}, satellite-to-ground quantum key distribution (QKD)~\cite{liaoSatellitetogroundQuantumKey2017,liaoSatelliteRelayedIntercontinentalQuantum2018}, ground-to-satellite quantum teleportation~\cite{ren_ground--satellite_2017}, and entanglement-based QKD over 1120~km~\cite{yinEntanglementbasedSecureQuantum2020}.
Since then, a plethora of other quantum-enabled satellite missions have been in various stages of development~\cite{reviewsats} including Jinan-1, which recently demonstrated real-time QKD using a weak coherent pulse source in combination with a classical optical channel for post-processing during the overpass~\cite{liMicrosatellitebasedRealtimeQuantum2024}.

Most satellite QKD missions are based on prepare-and-measure protocols, with a photon source on the satellite preparing quantum states and an optical ground station (OGS) measuring the quantum states---establishing a point-to-point down-link between two nodes. 
With the satellite acting as a trusted messenger, keys can be securely exchanged between distant OGSs, which are typically in remote locations.
We are instead interested in an up-link based satellite-quantum-communication scenario, where the ground-based quantum transmitter is an entangled photon source (EPS).
The OGSs are now untrusted nodes, supplying entangled photons to end-users within fibre connected networks and overpassing satellites simultaneously, a necessary building block in future quantum network infrastructure beyond point-to-point links~\cite{tubioSatelliteassistedQuantumCommunication2024}.
The typical argument against up-link-based QKD is the higher channel loss compared to down-link, with a total loss of between 41 and 52~dB to be overcome~\cite{ren_ground--satellite_2017}, which requires high-performance and low-noise quantum light sources.

In this work, we present an ultra-bright EPS with far non-degenerate wavelengths suitable for simultaneous transmission through an up-link to a LEO satellite and a fibre network.
We emulate quantum key acquisition for up-link loss profiles typical of a LEO satellite such as QEYSSat~\cite{hasan_qeyssat_2014,srikara_quantum_2024}, with a simultaneous fibre link long enough to connect to an end-user remote network.
This type of emulation has been performed for decoy-state QKD \cite{shields,mendes_optical_2024}.

\begin{figure}[ht!]
\begin{center}
\includegraphics[width=.85\columnwidth]{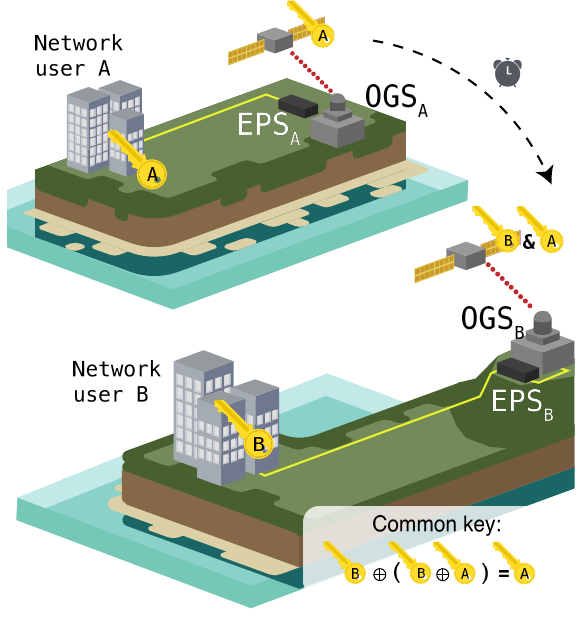}
\end{center}
\caption{\textbf{Schematic for conducting up-link QKD with an overpassing satellite.} 
    The EPS located at the OGS distributes entangled photons between an overpassing satellite and an end-user located in a network, Alice, allowing them to generate a secret key $k_{\text{A}}$.
    The satellite, a trusted node, then travels to another end-user, Bob, and they similarly generate $k_{\text{B}}$.
    The satellite now uses $k_{\text{B}}$ as a one-time-pad to encode $k_{\text{A}}$ and securely share it with Bob.
    }\label{fig:overview}
\end{figure}

Figure~\ref{fig:overview} shows a scenario for global quantum key distribution with photonic entanglement serving local networks while connecting to satellites via up-links. 
We operate the up-link at 785~nm which is found to be optimal for this purpose by trading off transmission loss, diffraction loss and efficiency of space-qualifiable single photon detectors~\cite{hasan_qeyssat_2014,bourgoin_comprehensive_2013}, and is inside the wavelength range of QEYSSat's receiver. 
For low-loss fibre transmission, the second photon should ideally be in the telecom C-band~\cite{oh_polarization_2024,valivarthi}.

To generate entangled photons at these wavelengths, a highly non-degenerate parametric down-conversion (PDC) process can be used~\cite{peltonBrightSinglespatialmodeSource2004}. 
The first use case for this type of EPS, at a time when high-efficiency telecom single-photon detectors were not yet available, was to use the $\sim$~780~nm photon as a herald for the 1550~nm photon~\cite{hubel2007high}.
Several other examples of highly non-degenerate photon sources demonstrate a strong interest in entanglement between different wavelengths for compatibility in a range of quantum tasks~\cite{Periscope, szlachetka_ultrabright_2023,konigEfficientSpectrallyBright2005,saugeSinglecrystalSourcePathpolarization2008,fiorentinoCompactSourcesPolarizationentangled2008, szlachetka_ultrabright_2023,oh_polarization_2024,zeitlerEntanglementVerificationHyperentangled2022,li_quantum_2024-1}.
A major challenge in scenarios involving two far non-degenerate wavelengths is to achieve a high degree of accuracy in matching path-lengths, group delays, and spatial mode-overlap to attain high-quality entanglement.

\begin{figure}[ht!]
\begin{center}
\includegraphics[width=0.8\columnwidth]{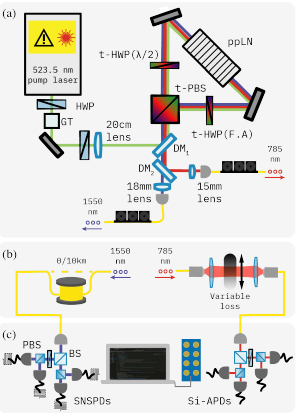}
\vspace{-1em}
\end{center}
\caption{\textbf{Experimental schematic.} 
    (a) Entangled photon source (description in main text).
    (b) The telecom photon is transmitted through 10~km of fibre whilst the 785~nm photon is transmitted through a free-space channel.
    The free-space channel contains a variable loss wheel, which is mounted to a motorised controller to emulate the loss profile of an overpassing satellite.
    (c) After the fibre and free-space channel transmission respectively, the entangled photons are measured in either the Pauli-$Z$ or Pauli-$X$ bases to perform the BBM92 quantum communication protocol.
    A random basis choice is made by a balanced beam-splitter (BS) and photons are projected into the Pauli-$Z$ basis using a polarising beam-splitter (PBS) or projected into the Pauli-$X$ basis using a half-wave-plate (HWP) and a PBS.
    After projection, the photons are again coupled into single-mode fibres, the telecom photons are detected by superconducting nanowire single-photon detectors (SNPDs) and the $785$~nm photons are detected using Silicon single photon avalanche detectors (Si-APDs).
    Detection events are tagged with a time stamp by a time-tagging unit, and the tags are subsequently processed to identify detection events that correspond to coincidence events.}\label{fig:source}
\end{figure}

The entangled photon source used in our emulation is based on photon pair creation via PDC in a 20~mm long, 5~\% MgO doped, periodically-poled Lithium Niobate (ppLN) crystal (from Covesion Ltd), with a poling period of 7.5~{\textmu}m, phase-matched to create vertically co-polarised photons at 1572 nm and 785 nm from 523.5~nm pump photons at a temperature of 84°C. 
Type-0 was chosen because it offers the highest nonlinear susceptibility coefficient and thus the highest available brightness. 
The crystal is embedded in a Sagnac interferometer~\cite{fedrizzi_wavelength-tunable_2007,kimPhasestableSourcePolarizationentangled2006}, as shown in Figure~\ref{fig:source}.
Diagonally polarised pump photons from a continuous wave diode-pumped solid state laser (Frankfurt Laser Company) are focused with a 20~cm lens before passing through a tri-chromatic polarising beam splitter (t-PBS) specified for all three system wavelengths. 
A key component in Sagnac-type photon sources is a custom half-wave plate (t-HWP) which rotates the pump beam by $\pi/2$ while also flipping the counter-propagating PDC photons by $\pi/2$, such that photon pairs from both directions exit the PBS through the original input port.
This component can be designed with high optical quality for two wavelengths, but not necessarily for three.
(Super-)achromatic retarders are typically highly multi-order and have, in our tests, proven to be of inferior optical quality, preventing accurate spatial mode matching.
Instead, we use a low-order wave-plate designed by Bernard-Halle which approximates optimal retardance to within reasonable values: 0.02~$\lambda/2$ for 523.5~nm, 0.07~$\lambda/2$ for 785~nm, 0.02$\lambda/2$ for 1572~nm.
To ensure that any temporal offset by the HWP between the two down-converted photons is experienced in both directions around the Sagnac and retain temporal coherence in the Bell state, we place an identical HWP at its fast axis on the other side of the crystal.
Finally, the photon pairs are separated from the pump beam using a long pass dichroic mirror, and then from each other using a second dichroic mirror.
Both are designed by Envin Scientific to have minimal polarisation dependence.
The pump waist (radius) is $\sim$~33~{\textmu}m.
The 1572~nm photons are captured with collection waist of 90~{\textmu}m and detected  with superconducting nanowire single photon detectors with a nominal detection efficiency of $\sim$~80~\%, dark count rate of up to 100~cps, temporal jitter of 50~ps and a dead time of 25~ns.
The 785~nm photons are captured with a collection waist of 65~{\textmu}m and detected by single photon avalanche detectors (Si-APDs) which operate with a quantum efficiency of 60~\%, dark count rate $>$50~cps, temporal jitter of 350~ps and a maximum dead time of 30~ns.
The entangled photons have bandwidths of 0.8~nm and 2.3~nm for the $\sim$~785~nm and $\sim$~1572~nm photon respectively (theoretically calculated using Ref.~\cite{spdcalc}), and we use a 3~nm bandpass filter for the $\sim$~785~nm photon to filter our any unwanted noise.
The detector signals are recorded by a QuTools multi-channel time-tagger with 1~ps resolution and a nominal jitter of 6~ps. 
Coincident detection events are established in post-processing within a coincidence timing window (t$_{CC}$) using locally developed software.
We obtained a maximum measured brightness of 223~kHz/mW whilst achieving a (symmetric-)heralding efficiency of 19.3~\%, calculated using $CC/\sqrt{s_{1}s_{2}}$ where $s_1$ and $s_2$ are the singles detected in the signal and idler and $CC$ are the coincidences between them.
With up to 1.6~W of pump power available, we can in principle create 660~MHz detectable photon pairs (assuming sufficiently fast detectors).
For comparison, the reported measured brightness from other recent highly non-degenerate EPSs were 33~kHz/mW~\cite{oh_polarization_2024} with a symmetric heralding efficiency of 10.7~\% and 70~kHz/mW~\cite{szlachetka_ultrabright_2023} with a symmetric heralding efficiency of 10.5~\%. 

The EPS creates states of the form 
\begin{equation} 
\ket{\Phi^+}=\cos{\alpha}\ket{H_{785}H_{1572}}+\sin{\alpha}e^{i\phi}\ket{V_{785}V_{1572}},
\end{equation}
where the angle $\alpha$ and phase $\phi$ are controlled by the polarisation state of the pump beam.
To characterise the entangled state quality, we perform full quantum state tomography by projecting the two photons onto all 36 combinations of the 6 polarisation states ${\ket{H},\ket{V},1/\sqrt{2}(\ket{H}\pm\ket{V}),1/\sqrt{2}(\ket{H}\pm i\ket{V})}$ using a half-wave plate (HWP), quarter-wave plate (QWP), and a polarising beam splitter (PBS). 
From these measurements, we reconstruct a density operator $\rho$ via maximum likelihood estimation, and estimate an entangled state purity of $\mathcal{P}=\Tr(\rho^2)=95.6\pm0.4$\%, where the error is calculated based on 1000 Monte Carlo iterations sampling from a Poissonian photon number distribution.
The main factor limiting state purity is the tri-chromatic HWP applying non-ideal retardation.

\begin{figure}[ht!]
\begin{center}
\includegraphics[width=0.95\columnwidth]{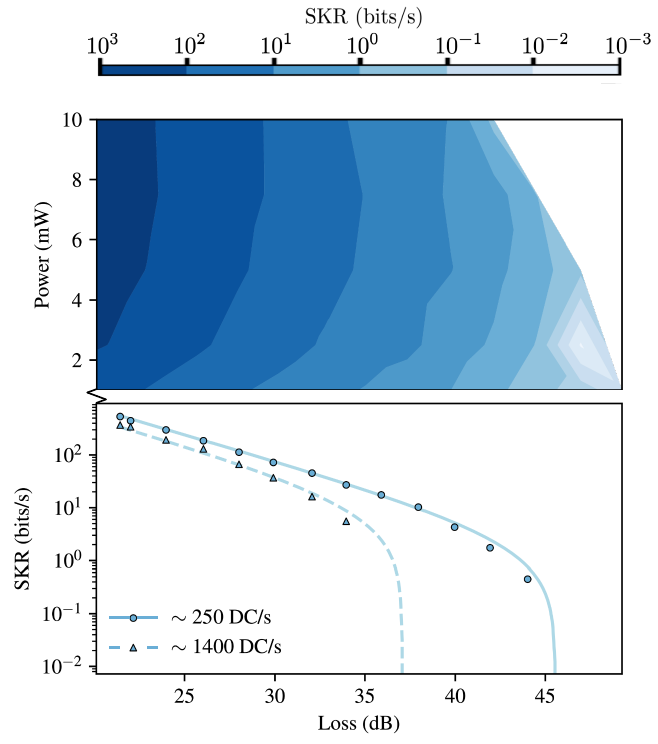}
\end{center}\vspace{-1em}
\caption{
    ~\textbf{Secure key rate estimates with 10~km fibre.} 
    Top: SKR in the asymptotic limit as a function of satellite up-link loss, with a fixed coincidence window for different pump powers.
    Bottom: SKR in the asymptotic limit versus link loss for two different Si-APD dark count rates. The SKR is non-zero up to $\sim$~35~dB with a dark count rate of 250~cps, and up to $\sim$~45~dB with a dark count rate of 1400~cps.
    The solid and dashed lines are the simulated results using the model contained in Ref.~\cite{neumann} for our experimental parameters.
    Data points are evaluated via Eq.~(\ref{eq:SKR}).
}\label{fig:loss_shoulder}
\end{figure}
\begin{figure}[t!]
\begin{center}
\includegraphics[width=.9\columnwidth]{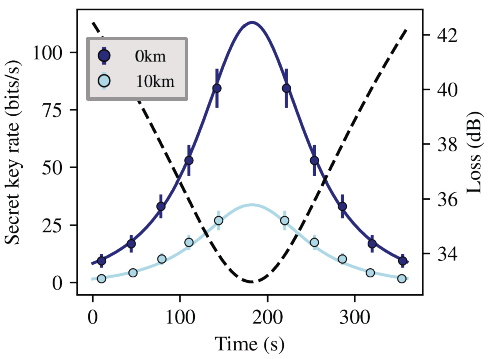}
\end{center}\vspace{-1em}
\caption{
    \textbf{Experimental up-link emulation.} 
    To determine the anticipated key length from an overpass via an up-link channel we extract the empirical channel loss data from Ref.~\cite{ren_ground--satellite_2017}, plotted as the black line.
    Combining our results, the modelling work in Ref.~\cite{neumann} and the extracted channel loss, the purple and dark blue lines show how the SKR varies as a function of time for the scenarios with 0~km and 10~km of telecom fibre spool attached, at a fixed pump power of 5~mW.
    Without terrestrial fibre, the asymptotic key rate peaks at 113~bps creating an integrated key of 18146~bits in one overpass.
    When the terrestrial photons travel through 10~km of fibre, the SKR peaks at 33.9~bps and integrates to a key of 5223~bits in one overpass.
}\label{fig:simulation}
\end{figure}

Having introduced our EPS we now proceed to emulating the BBM92 quantum communication protocol~\cite{bennett_quantum_1992} in the high-loss satellite up-link scenario shown in Figure~\ref{fig:overview}. 
Alice and Bob receive one half each of an entangled photon pair distributed by the OGS. 
Each of their receivers measures in either the Pauli-$X$ or $Z$ bases.
Once both parties have performed a number of measurement rounds, they publicly announce their measurement bases and discard the measurements made in uncorrelated bases. 
Whenever they measured in the same basis, the entanglement ensures that they received correlated (modulo experimental errors) random outcomes, i.e. key bit values. 
From this raw key, a secure key is then post-processed via error correction and privacy amplification.

To evaluate our up-link emulation we determine secure key rates (SKR) using the definition
\begin{equation}\label{eq:SKR}
    \text{SKR} = \frac{CC}{2} (1 - h(\text{QBER}) - h(\text{Q$_{\text{X}}$})),
\end{equation}
where $h$ is the binary entropy function.
The quantum bit error rate (QBER) is defined for the $\ket{\Phi^+}$ state as the probability that a $Z$ basis coincidence detection is anti-correlated:
\begin{equation}
    \text{QBER} = 1-\frac{\sum_{i\in\{H,V\}}<ii>}{\sum_{i,j\in\{H,V\}}<ij>}.
\end{equation}
Finally, Qx is the phase error, which has an equivalent definition to the QBER, but for the Pauli-$X$ basis.
At zero distance, we measured a QBER of 1.7~\% and Qx of 3.9~\%, producing an SKR of 141.57~kbps at a pump power of 1~mW.
These values will be drastically lower during operation of a satellite up-link because of the substantial amount of loss the photons will experience before detection.
Free-space links through the atmosphere have been modelled extensively~\cite{peng_experimental_2005,hasan_qeyssat_2014,sidhu_finite_2022}.
The primary factors contributing to link loss are beam diffraction from the limited transmitter telescope aperture, beam wander caused by atmospheric turbulence, and the limited collection angle from the receiver telescope. 
The loss profile we use in this work is extracted from Ref.~\cite{ren_ground--satellite_2017}, which showed empirically measured total losses between 52~dB and 41~dB at elevations of 14.5° and 76°, respectively, during an up-link from an OGS and a LEO satellite equipped with a receiver payload.
We emulate this loss profile using a short free-space link equipped with a fixed 20~dB neutral density (ND) filter plus a variable 0.4~-~40~dB ND filter wheel mounted onto a motorised rotation stage to introduce controlled loss to the 785~nm channel.
To represent the terrestrial link between a metropolitan end-user and the OGS, the 1572~nm photons are sent through 10~km of standard telecom fibre. 

Before emulating the overpass of a satellite, we determine the maximum SKR our EPS source can produce, as well as the maximum tolerable loss our EPS source can sustain for a range of different pump powers. 
We vary the amount of loss in the system via the variable loss wheel in the 785~nm channel.
This loss is on top of the already present fixed losses from non-ideal optics, coupling inefficiencies, as well as a contribution from the 10~km of fibre in the 1572~nm channel.
These results are shown in Figure~\ref{fig:loss_shoulder}.
To obtain a statistically significant result for the SKR, we collect at least 100 coincidences per basis. 
The physical effects that contribute to the experimentally observed SKR, such as the Poissonian nature of PDC, jitter from several sources, dark counts and dead time, have been captured in a simulation by Neumann et al.~\cite{neumann} which we use to validate the experimentally measured data.
As expected, the SKR decreases exponentially as loss increases, with an eventual cut-off at the point at which the detected signal count rates are of the same order as the detector dark counts, causing the QBER and Qx to rapidly increase beyond a level where there is a positive SKR.
This is more noticeable in the bottom pane of Figure~\ref{fig:loss_shoulder}, which also shows the SKR for dark count rates typical in laboratory environments, as well as higher dark count rates that are more representative of satellite deployed APDs.

Knowing the source performance for a variety of powers and under a range of losses, we can now quantify the SKR from an emulated overpass, shown in Figure~\ref{fig:simulation}.
We emulate an overpass by extracting and applying the loss profile from Ref.~\cite{ren_ground--satellite_2017} (which provides an empirically measured loss profile for an up-link ground-to-satellite channel). 
For the case with 0~km and 10~km of terrestrial fibre we use a coincidence window of 400~ps or 1~ns respectively.
Since our experimental data already incorporates detector inefficiency of the Si-APDs, we subtract the 3~dB that these contributed to the loss profile. 
Following the same argument, we subtract a further 5.9~dB that were attributed to optical losses (again see Ref.~\cite{ren_ground--satellite_2017}) which our experimental data also already incorporates. 
These deducted losses are considered in more detail within the Supplementary Materials.
In the scenario without a terrestrial fibre link, the SKR peaks at 113~bps and a key of 18146~bits in an emulated overpass is obtained.
In the scenario where the telecom photons pass through 10~km of fibre, the SKR peaks at 33.9~bps and we accumulate a key of 5223~bits per overpass. These numbers indicate that despite the challenging up-link loss profile, a significant amount of secure key will be obtainable at least in the asymptotic limit. 
It also suggests that we would be able to recover a finite key \cite{lim_security_2021}.

We have identified several means to increase the SKR.
Improving the quality of the quantum state is a limited avenue for improving the system performance.
Modelling how the SKR depends on state fidelity reveals a perfect state fidelity would gain us 1.9~dB of loss budget, see Supplementary Materials. Over the course of an overpass this would increase the secure key length by 4~kbit.
Improved heralding is a more straightforward means of increasing the loss budget since more singles will constitute true coincidences rather than contribute to the accidental coincidences involving dark counts.
With collection efficiencies obtained in previous works~\cite{Pickston:21}, with our detectors we should be able to roughly double the heralding efficiency. This would increase the loss budget by $\sim$~4~dB in our work or increase the secure key length in an overpass by 26~kbit. 
To achieve this greater heralding efficiency we can use a looser focusing condition, this is our main focus during further development of the EPS.

The fibre distance for the terrestrial link is currently limited by chromatic dispersion, which beyond 10~km broadens our photons beyond the inherent temporal detection jitter. This prevents the coincidence window from being optimised for maximum signal-to-noise. 
Dispersion compensators can undo this effect but they typically add $\sim$~5~dB of insertion loss. The Supplementary Material contains detailed modelling on the fibre link.
Adaptive optics can be used to reduce ground-to-satellite channel loss by 3~-~7~dB~\cite{pugh_adaptive_2020}, increasing the SKR by countering the effects of turbulence.

One of the things that most drastically affects the loss budget is the dark counts.
The levels we've chosen are representative for contemporary missions like QEYSSat \cite{background_meas} which operate with APDs on board, but progress in detector technology~\cite{yangSpaceborneLownoiseSinglephoton2019} promises to bring down the intrinsic dark counts.
The choice of the OGS location also plays an important role, as shown in~\cite{yastremski2024estimatingimpactlightpollution}.

The coincidence window, t$_{CC}$, is subject to a trade-off that can be optimised, as shown in Figure~\ref{fig:tcc_skr} of the Supplementary Materials.
A lowered t$_{CC}$ reduces the noise parameters as fewer dark counts will be accidentally identified as coincidences, and fewer separate down-conversion events will occur within t$_{CC}$ of each other, effectively acting as a time gate and allowing to pump harder.
On the other hand it comes with a disadvantage to the raw coincidence rate once it is on the order of the system jitter, as true coincidences start being rejected.

The system jitter in our setup is dominated by the up-link APDs.
We expect that in a deployment scenario, where the two photons of each pair are registered on separate timetaggers, the synchronisation jitter will be another significant contribution to the overall jitter.
This arises because the detection events are recorded relative to two imperfectly synchronised clocks.
There are methods to enhance clock synchronisation and reduce synchronisation jitter---some proposals require similar infrastructure to that being used in this work~\cite{dai_towards_2020,haldar_towards_2023}.

Based on the distribution of entanglement, one can envision how this work fits into a future quantum internet consisting of terrestrial and satellite nodes. 
Firstly, if the satellite were to contain a Bell state measurement (BSM)-based receiver---as proposed in Ref.~\cite{srikara_quantum_2024}--then end-users could be directly entangled and each node will become untrusted.  
However, to achieve this, one would need to operate the source in the pulsed regime, and the photons created by the EPS need to be engineered to maximise two-photon interference visibilities.
This can be achieved in PDC by using domain-engineering techniques~\cite{graffitti_pure_2017,pickstonOptimisedDomainengineeredCrystals2021}.
Additionally, by performing entangling operations~\cite{booneEntanglementGlobalDistances2015} one could demonstrate more complex quantum information protocols, such as a multi-user quantum communication protocols like quantum conference key agreement~\cite{murtaQuantumConferenceKey2020,originalQCKA, pickstonExperimentalNetworkAdvantage2022}.
Since we only use a fraction of our laser's pump power, our EPS could be multiplexed to serve several pairs of end-users simultaneously~\cite{vinet_reconfigurable_2025}.
Finally, one could explore creating far non-degenerate entanglement encoded in frequency bins~\cite{morrisonFrequencybinEntanglementDomainengineered2022}, or pulse modes for multiplexing~\cite{graffitti_direct_2020}.

Note added: while preparing this manuscript, we became aware of a similar independent work whose findings are consistent with ours~\cite{joarder2025entanglementbasedquantumkeydistribution}. 

\section*{Acknowledgements}
 We would like to thank J. Ho, P. Barrow and C. L. Morrison for assistance. This work was supported by the UK Engineering and Physical Sciences Research Council (Grant Nos, EP/T001011/1, EP/Z533208/1 and EP/W003252/1.) 

\bibliography{bib}
\bibliographystyle{naturemag}

\clearpage
\setcounter{figure}{0} 
\onecolumngrid
\section*{Supplementary material}

\renewcommand{\thefigure}{S\arabic{figure}}
\setcounter{figure}{0}

\subsection{Dispersion in telecom fibre}\label{supp1}
\begin{figure}[h!]
    \centering
    \includegraphics[width=0.9\columnwidth]{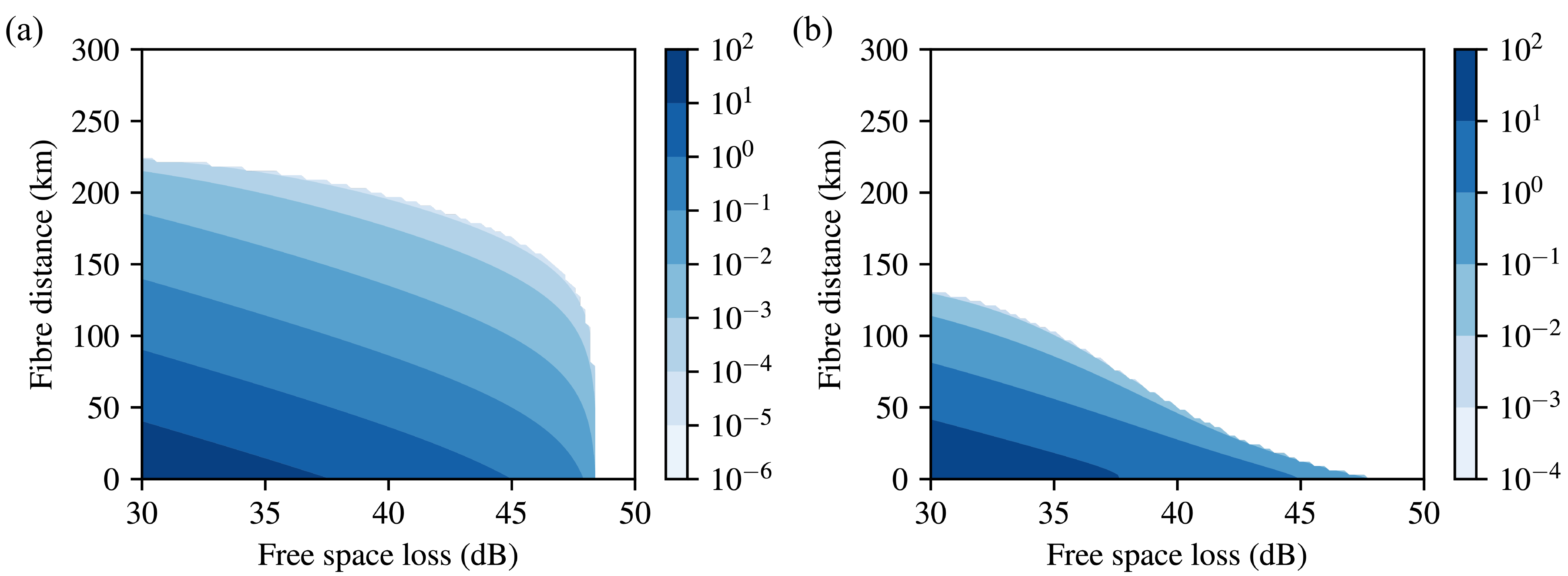}
    \caption{(a) Contour plot of the theoretical SKR given our system parameters assuming no dispersion.
    (b) Contour plot of the theoretical SKR given our system parameters assuming a dispersion of 18~ps/km.}
    \label{fig:disp_no_disp}
\end{figure}

\vspace{-.5em}
The telecom photons we produce in our EPS are transmitted through SMF-28. 
As we generate photons centred at $\sim$ 1572~nm, the dispersion coefficient at this wavelength is greater than 18~\text{ps/km/nm} for photons centred at 1550~nm. 
With this in mind, and given a photon bandwidth of 2.2~\text{nm} (calculated using SPDCalc.~\cite{spdcalc}), the dispersion sustained is at least 40~ps/km, therefore after 10~km, 400~ps of dispersion is accumulated.
This is on the order of the system jitter, thus greater fibre distances will become the dominant contribution to jitter and require an increased coincidence window.
Increasing the coincidence window is possible, but as we noted in the main text, will result in a lower signal to noise ratio.
A dispersion compensator can mitigate the fibre's effective contribution to jitter, and non-local dispersion compensation~\cite{neumann_experimentally_2021,franson_nonlocal_1992} means that the application of negative dispersion is only required on one of the entangled photons. 
Figure~\ref{fig:disp_no_disp} shows the effect of dispersion on the attainable SKR as a function of free space loss and distance in fibre.

\subsection{Effect of state fidelity and heralding efficiency on SKR}\label{supp2}
\begin{figure*}[h!]
    \centering
    \includegraphics[width=.4\columnwidth]{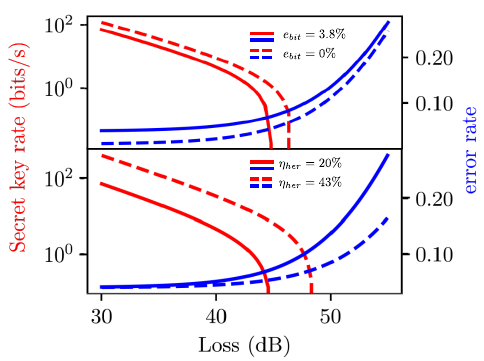}
    \caption{The loss budget increase for perfect state fidelity would be 1.9~dB, while a heralding efficiency of 43\% would increase the loss budget by 4~dB.}
    \label{fig:supp_fig_2}
\end{figure*}

\vspace{-.5em}
To influence design considerations for a future iteration of this source we investigated how the source's state fidelity and heralding efficiency effects the SKR. 
Our evaluation focused on how improvements would change the amount of tolerable loss (and as a consequence produce the highest SKR).
For perfect state fidelity, the tolerable loss would increase by 1.9~dB, whilst a doubling in heralding efficiency would increase the the tolerable loss by 4~dB, as shown in Figure~\ref{fig:supp_fig_2}.
Our entangled photon source is able to produce a pair rate far greater than the optimal rate for our scenario.
Therefore, in the next iteration we will sacrifice brightness for increased heralding efficiency by changing the focusing condition to produce a larger spot size~\cite{pickstonOptimisedDomainengineeredCrystals2021}.

\subsection{Overpass loss profiles}
\begin{figure}[h!]
    \centering
    \includegraphics[width=.4\columnwidth]{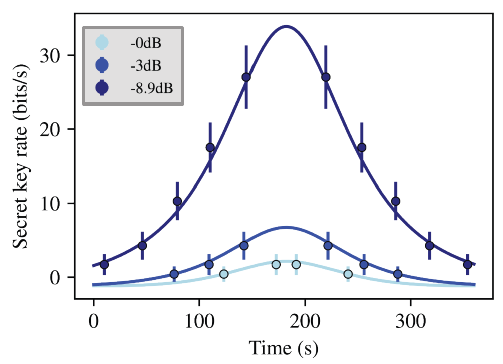}
    \caption{SKR during a satellite overpass
     with the loss profile from Ref.~\cite{ren_ground--satellite_2017}, with 3 different correction factors. During an uncorrected overpass, the SKR peaks at 2.2~bps and integrates to a key of 178~bits.
     When we correct for the double counted detector inefficiencies, the SKR peaks at 9.5~bps and integrates to a key of 782~bits.
     When we additionally correct for the optical losses, the SKR peaks at 33.9~bps and integrates to a key of 5223~bits.}
    \label{fig:different_levels}
\end{figure}

\vspace{-.5em}
In the main text, Figure~\ref{fig:loss_shoulder} uses a loss profile extracted from Ref.~\cite{ren_ground--satellite_2017}.
As we wanted just the channel loss, we deducted 8.9~dB from the total loss profile, which accounted for 3~dB of loss from non-unit detector quantum efficiency, as well as 5.9~dB of optical losses in their system---these values are quoted in Ref.~\cite{ren_ground--satellite_2017}.
We show in Figure~\ref{fig:different_levels} the overpass with the uncorrected overpass, as well as when correcting for the detector efficiency, and when additionally correcting for optical losses, showing that even in a scenario of more loss, we still attain positive key rates.

\subsection{Real-time optimisation of source and data collection}
\begin{figure}[h!]
    \centering
    \includegraphics[width=.4\columnwidth]{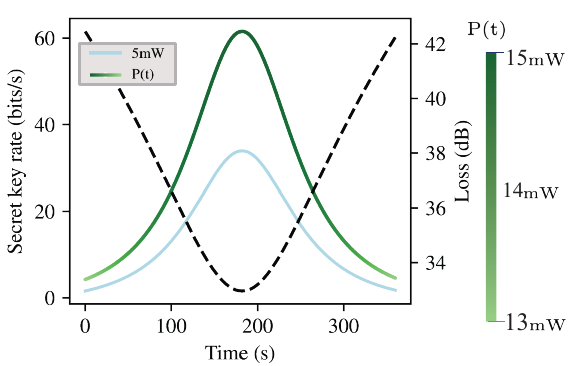}
    \caption{Optimised SKR during a satellite overpass
    , corrected for the optical losses and detector inefficiencies measured in Ref.~\cite{ren_ground--satellite_2017}.}
    \label{fig:-8.9db_overpass}
\end{figure}

\begin{figure}[t!]
    \begin{center}
    \includegraphics[width=.4\columnwidth]{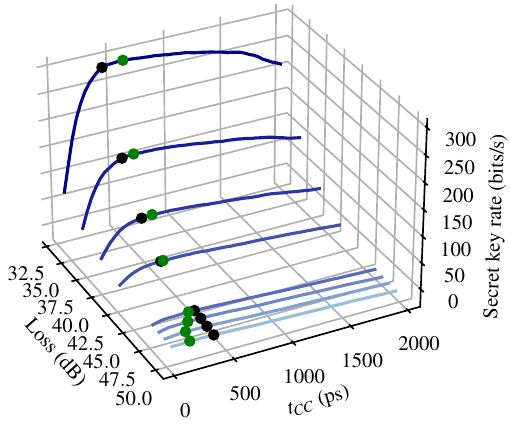}
    \end{center}\vspace{-1em}
    \caption{Effect of instantaneous optimisation of pump power and coincidence window on the SKR as a function of both loss, and the $t_{CC}$.
    For each loss value in this plot, we collected a set of timetags using 1~mW of pump power and extract the SKR using a fixed $t_{CC}$, shown as the black data points on this curve. 
    We also show the optimal $t_{CC}$ in green, by re-processing the tags for a range of different $t_{CC}$ values between 0 and 2~ns. }
    \label{fig:tcc_skr}
\end{figure}

\vspace{-.5em}
One thing we considered, is how to optimise the SKR during an overpass. 
We know that there is an optimal source brightness (which we control via the amount of pump power used) for each value of loss. 
As the loss changes during an overpass, we wanted to analyse how changing the amount of pump power during this time would provide more keys.
In addition to optimal source brightness, there is also an optimal coincidence window, which again changes as a function of loss, and thus also as a function of time during an overpass.
Figure~\ref{fig:-8.9db_overpass} compares the secret key rate during an overpass using a fixed pump power of 5~mW to the secret key rate generated if we were to use an instantaneously and continuously optimised pump power and coincidence window over the duration of the overpass.
For this optimised scenario, we report SKR peaking at 61.5~bps and creating a key length of 9768~bits, an increase of $\sim$~78~\%.
In addition, Figure~\ref{fig:tcc_skr} shows how the optimal coincidence window (green dots) changes as the loss profile changes. 

\clearpage
\newpage

\end{document}

%% file: formatting.tex
\usepackage[utf8]{inputenc}
\usepackage[T1]{fontenc}
\usepackage{braket}
\usepackage{graphicx}
\usepackage{booktabs}
\usepackage{mathtools}	
\usepackage{amssymb}
\usepackage{amsmath}	
\usepackage{amsfonts}
\usepackage{mathrsfs}
\usepackage{color}
\usepackage{upgreek}
\usepackage{amsthm}
\usepackage{psfrag}
\usepackage{ifsym}
\usepackage{dsfont}
\usepackage{enumerate}
\usepackage{enumitem}
\usepackage{multirow} 
\usepackage{lipsum}
\usepackage{float}
\usepackage[separate-uncertainty=true]{siunitx}
\usepackage{setspace}
\usepackage[dvipsnames]{xcolor}
\usepackage[normalem]{ulem}
\usepackage{textgreek}

\usepackage{silence}
\usepackage{multirow}
\usepackage{booktabs}

\usepackage{hyperref}
\definecolor{refColour}{RGB}{0, 0, 0}
\definecolor{citeColour}{RGB}{0, 0, 0}
\definecolor{extraColour}{RGB}{0, 0, 0}
\hypersetup{colorlinks=true,linkcolor=Magenta,citecolor=Magenta}

\usepackage{titlesec}
\titleformat*{\subsection}{\normalfont\centering}
\titleformat*{\section}{\normalfont\centering\large}

\newcommand{\Tr} {\operatorname{Tr}}

\newcommand{\one}{\leavevmode\hbox{\small1\normalsize\kern-.33em1}}